# Long-term Stability of Bilayer $MoS_2$ in Ambient Air

John Femi-Oyetoro[1], Kevin Yao[1, a], Evan Hathaway[1], Yan Jiang[1], Ibikunle Ojo[1], Brian Squires[1], Arup Neogi[1], Jingbiao Cui[1], Usha Philipose[1], Nithish Kumar Gadiyaram[2], Weidong Zhou[2], and Jose Perez[1, *]

[1]Department of Physics, University of North Texas, Denton, TX 76203, United States

[2]Department of Electrical Engineering, University of Texas at Arlington, Arlington, Texas 76019, United States

**Abstract**

We report that chemical vapor deposition (CVD) grown bilayer and thicker-layered $MoS_2$ are structurally and optically stable under ambient conditions, in comparison to CVD-grown monolayer $MoS_2$ and other transition metal dichalcogenides (TMDs) that have been reported to degrade under the same conditions, hindering their many potential applications. We present atomic force microscopy (AFM), and Raman and photoluminescence (PL) spectroscopy measurements of as-grown and preheated multilayer $MoS_2$ after exposure to ambient conditions for periods of up to 2 years. The AFM images show that, under ambient conditions, as-grown and preheated bilayer and thicker-layered $MoS_2$ films do not exhibit the growth of dendrites that is characteristic of monolayer degradation. Dendrites are observed to stop at the monolayer-bilayer boundary. Raman and PL spectra of the aged bilayer and thicker-layered films are comparable to those of as-grown films. The greater stability of bilayers and thicker layers supports a previously reported mechanism for monolayer degradation involving Förster resonance energy transfer. Our results show that bilayer and thicker-layered TMDs are promising materials for applications requiring ambient stability.

*Corresponding author.
E-mail address: jperez@unt.edu (J.M. Perez).

[a]Current address: Texas A&M University, College Station, TX 77843, United States

## Introduction

Two-dimensional (2D) materials have attracted considerable interest due to their unique properties that have potential applications in a wide variety of areas. However, most of the discovered 2D materials quickly degrade under ambient conditions,[1] significantly impeding their use in practical devices. For example, monolayer (ML) $MoS_2$ is a transition metal dichalcogenide (TMD) that may have unique applications due to its direct bandgap and large spin-orbit coupling.[2] However, previous reports have shown that chemical vapor deposition (CVD)-grown ML $MoS_2$ and other ML TMDs degrade under ambient conditions within a year.[3–7] Despite the variety of degradation prevention techniques such as encapsulation with hexagonal boron nitride[8] and polymers[9], and placement in an environment with a desiccant[3] or in a vacuum,[5] these techniques are not feasible for large-scale applications. To our knowledge, the stability of bilayer (BL) and thicker-layered $MoS_2$ under ambient conditions has not been extensively studied. In this paper, we report on the stability of as-grown and preheated BL and thicker-layered $MoS_2$. The samples are grown using CVD on $SiO_2$ substrates and studied using atomic force microscopy (AFM), and Raman and photoluminescence (PL) spectroscopies. Samples are preheated to accelerate the degradation process in order to study it in a practical short time span. We find that BL and thicker-layered samples are remarkably stable under ambient conditions.

Bilayer and thicker-layered $MoS_2$ films, although possessing an indirect bandgap, have attracted considerable interest because of useful electronic properties not possessed by $MoS_2$ MLs. These properties give BLs and thicker layers certain advantages over MLs in the fabrication of electronic and photonic devices. For example, they have high electrical conductivity as compared to MLs due to their high density of states and effective screening of impurities in the substrate. Multilayer TMD transistors have been shown to have an excellent combination of high on/off current ratio and on-state current.[10] On the other hand, ML TMDs have a relatively small on-state current despite possessing greater on/off current ratios. In addition,

BL device yield is typically much higher than ML device yield, due to the greater mechanical stiffness of BLs.[11,12] Finally, BL and thicker-layered TMDs offer control over properties such as spin-orbit coupling,[13] interlayer coupling,[14] and band gap.[15] This ability to precisely tune the physical and electronic properties of atomically thin TMDs have potential applications in novel lateral heterostructures.[15–17] Also, varying the twist angle between layers in BLs can lead to new and potentially useful properties such as unconventional superconductivity at small twist angles,[18] twist dependent valley and band alignment,[19] and moiré pattern excitons.[20]

The ambient degradation of CVD-grown ML $MoS_2$ was first reported by Gao et al.[3] They observed that ML $MoS_2$ and $WS_2$ grown on $SiO_2$ substrates developed extensive cracking, morphological changes, and quenching of photoluminescence (PL) after exposure to an ambient environment at room temperature (RT) for a period of about a year. The degradation was attributed to oxidation along the grain boundaries and the adsorption of organic contaminants on the films. It was found that water vapor in the air was necessary for degradation to occur since samples did not degrade in a dry box. In addition, there have been reports of $MoS_2$ degradation at RT and high humidities.[5,21] For instance, Budania et al.[5] reported that mechanically exfoliated thin multilayer $MoS_2$ on $SiO_2$ developed speckles in air at a relative humidity (RH) of 60% over a period of about a year,[5] which was attributed to the high concentration of water molecules in the air. Sar et al.[7] studied the degradation process of CVD-grown ML $MoS_2$ grown using two different sample positions and found that face-down grown samples developed cracks and degraded significantly as compared to horizontally-grown samples. It was concluded that the degradation process was a result of sulfur vacancy defects and film tensile stress that were affected by the growth conditions. In addition, Kotsakidis et al.[4] reported that the degradation of ML $WS_2$ and other TMDs in air was a photoinduced process involving excitation across the band gap, since degradation was not observed when samples under ambient conditions were kept in a dark environment or excited at photon energies below the trion exciton binding energy.[4] It was proposed that the degradation mechanism involved photo-oxidation, likely Förster resonance energy transfer (FRET) and/or photocatalysis. In addition, we have recently reported a rapid ambient degradation of CVD-grown ML MoS$_2$ that occurs over a period of weeks instead of a year after the MLs are preheated in air at temperatures in the range 285 - 330 ºC.[6] Both rapid degradation and natural degradation (in which samples are not preheated) were observed to involve the formation of dendritic structures, and thus it was proposed that rapid and natural degradation involve a similar mechanism.[6] The rapid degradation of preheated samples was attributed to an increase in oxidized sites due to the preheating. Finally, Qiu et al.[22] reported the degradation of the electrical properties of exfoliated BL MoS$_2$ field effect transistors from exposure to ambient air. The authors showed that after vacuum annealing at 77 ºC the electrical properties of as-fabricated devices significantly improved; five minutes after re-exposure to an oxygen environment, the electrical properties degraded. The degradation was found to be reversible upon annealing and attributed to the physisorption of oxygen and water. The effects of long-term exposure on the stability of the MoS$_2$ was not reported. We note that the degradation of ML TMDs involves morphological changes that occur over a long time and are not reversible. Thus, it is of vital importance to investigate if BL and thicker-layer MoS$_2$ is susceptible to degradation.

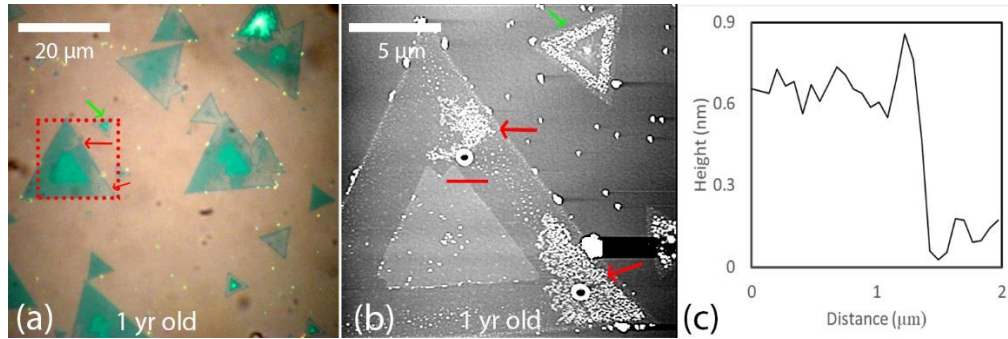

Figure 1. (a) Optical image of an MoS$_2$ monolayer and bilayer after about 1 year of exposure to ambient conditions. The arrows indicate areas of light optical contrast on the monolayer that are degraded. (b) AFM image of the boxed region in (a). (c) AFM height profile along the line in (b). The height of the inner MoS$_2$ island is about 0.6 nm corresponding to a bilayer.

**Results and Discussion**

Figure 1 (a) shows an optical image of an MoS$_2$ BL sample with an AA stacking sequence that was kept under ambient conditions for approximately one year after growth. In this paper, ambient conditions refer to air at RT and a RH of about 43%, and room lighting of about 1000 Lux that is typical of a lab. The image in Figure 1 (a) shows various BL MoS$_2$ islands with ML terraces. Figure 1 (b) is an AFM image of the boxed region in Figure 1 (a). Figure 1(c) shows an AFM height profile along the line in Figure 1 (b) showing a step height of about 0.6 nm consistent with that of a BL.[23] As indicated by the red arrows in Figure 1 (a), the ML terrace shows areas of light optical contrast that have been previously reported to be due to degradation.[3] Figure 1 (b) shows that the areas of light contrast contain dendrites. We note that the dendrites grow over the ML but are not present in the BL region. Also, it is noted that the small multilayer island on the top right-hand-side, indicated by the green arrow, has dendrites consuming the entire ML, but there are no such dendrites in the multilayer region of the island. This effect is widespread across the substrate. Almost all MLs on the 1.5-year-old substrate showed dendritic growths while BLs were unaffected, as shown in Figure S1. The degradation is irreversible, as shown in Figure S2, in which we vacuum annealed a sample at 500 °C for 2 h but the dendrites remained.

We also performed Raman and PL mapping of BL $MoS_2$ samples that had been under ambient conditions for approximately 1.5 years. We did not perform Raman or PL mapping immediately after growth to avoid effects of laser irradiation, which has been reported to degrade ML $WS_2$.[4] Figure 2 (a) shows an optical image of such a sample. The same regions of light optical contrast, which are attributed to degradation, are observed around the edges and growing into the ML. Various degraded areas in the ML are marked with numbers to assist in locating the areas in other figures. We note that there is an area of the ML that is adjacent to the BL, indicated by the number 1, that is degraded, but the adjacent BL appears unaffected. Figure 2 (b) shows a Raman map of the same sample depicting the relative height of the $E_{2g}$ peak at 382 cm$^{-1}$. The dark and light regions correspond to low and high peak heights, respectively. The $E_{2g}$ peak height from areas of the ML that have a light optical contrast in Figure 2 (a) is lower than that from areas of the ML that have a normal optical contrast. This effect has been previously attributed to the light-contrast areas being degraded.[3] However, the $E_{2g}$ and $A_{1g}$ peak height in the BL region appear uniform with no observable areas of reduced height that could indicate degradation. Figures S3 (a)-(f) show maps of the fitted peak height, position, and width of the $E_{2g}$ and $A_{1g}$ Raman peaks. This conclusion is further supported by the PL map of the A exciton peak centered at around 1.80 eV, shown in Figure 2 (c), which shows a spatially uniform height across the BL. The BL region exhibits a less intense PL as compared to the ML, as expected, due to BLs having an indirect band gap.[3] On the other hand, degraded ML regions exhibit decreased heights in the exciton peaks, consistent with previous reports.[3] In addition to an height decrease, we find that the A exciton peak of degraded MLs also shifts to higher energy by about 0.5 eV, as shown in Figure S4.

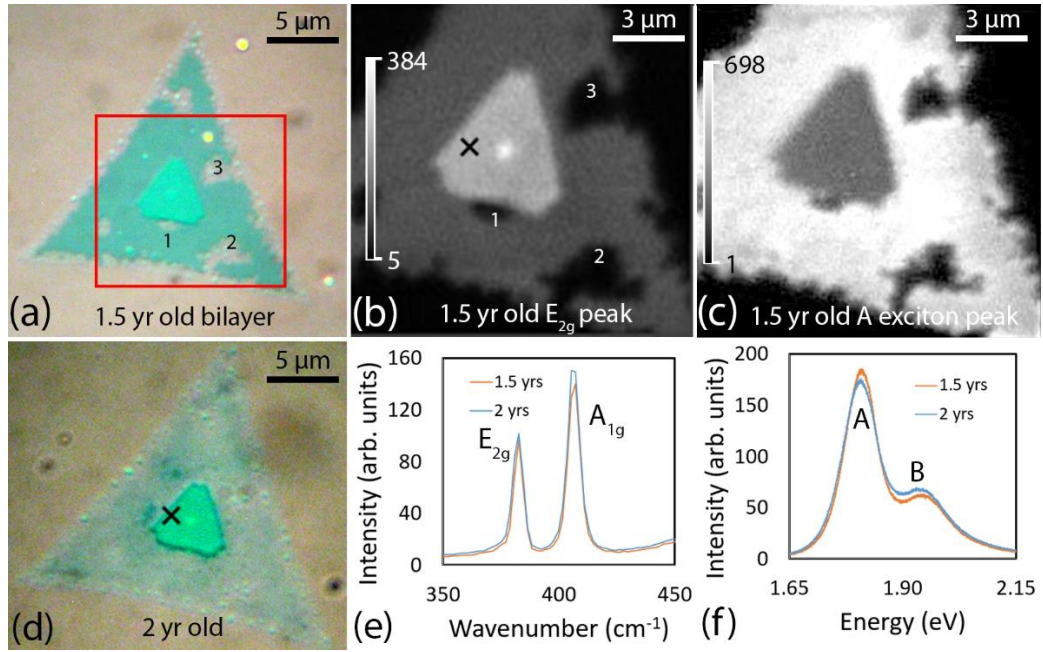

Figure 2. (a) Optical image of an MoS$_2$ monolayer with an AA-stacked bilayer after about 1.5 years of exposure to ambient conditions. The light optically contrasted areas on the monolayer, indicated by the numbers 1-3, show degradation. (b) and (c) Raman and photoluminescence maps of the area enclosed by the box in (a), showing the relative heights of the E$_{2g}$ and the A exciton peaks, respectively. (d) Optical image of the sample shown in (a) after 2.0 years of exposure to ambient conditions. The monolayer has almost completely degraded. (e) Raman spectra taken at the location of the cross in (b) and (d) after 1.5 and 2.0 years of exposure to ambient conditions. (e) Photoluminescence spectra taken at the location of the cross in (b) and (d) after 1.5 and 2.0 years of exposure to ambient conditions.

In order to determine if the Raman and PL spectra significantly change after additional exposure, we collected data on the same sample after an additional 0.5 years in ambient conditions for a total exposure of 2.0 years. Figure 2 (d) shows an optical image of the sample after 2.0 years showing that the ML has completely degraded but the BL still contains no observable regions of degradation. Figure 2 (e) shows representative Raman spectra of the BL taken at the location of the cross in Figure 2 (b) after 1.5 and 2.0 years. Figure 2 (e) shows representative PL spectra of the BL taken at the location of the cross in Figure 2 (b) after 1.5 and 2.0 years. The height of the PL spectra is normalized to their Si Raman peak at about 520 $cm^{-1}$. The fitted positions and full-width-at-half maxima (FWHM) of the Raman $E_{2g}$ and $A_{1g}$ peaks for both the 1.5 and 2.0-year-old samples are 382 $cm^{-1}$, 406 $cm^{-1}$, 4.9 $cm^{-1}$ and 6.5 $cm^{-1}$, respectively. The B and A exciton peaks for the 1.5-year-old sample are centered at, 1.95 eV and 1.80 eV, respectively, and the B and A exciton peaks for the 2-year-old sample are centered at 1.94 eV and 1.80 eV, respectively. The PL heights are also very comparable to each other. Our results for aged BLs are similar to previously published results on as-grown BLs: the center of the $E_{2g}$ peak and $A_{1g}$ peak of as-grown BLs have been reported to be around 381-382 $cm^{-1}$ and 404-405 $cm^{-1}$, respectively,[24] and the FWHM has been reported to be around 4.8-5.8 $cm^{-1}$ for the $E_{2g}$ peak and 5.8-6.7 $cm^{-1}$ for the $A_{1g}$ peak.[25] We conclude that the Raman and PL spectra of the BL has not undergone significant change during this time.

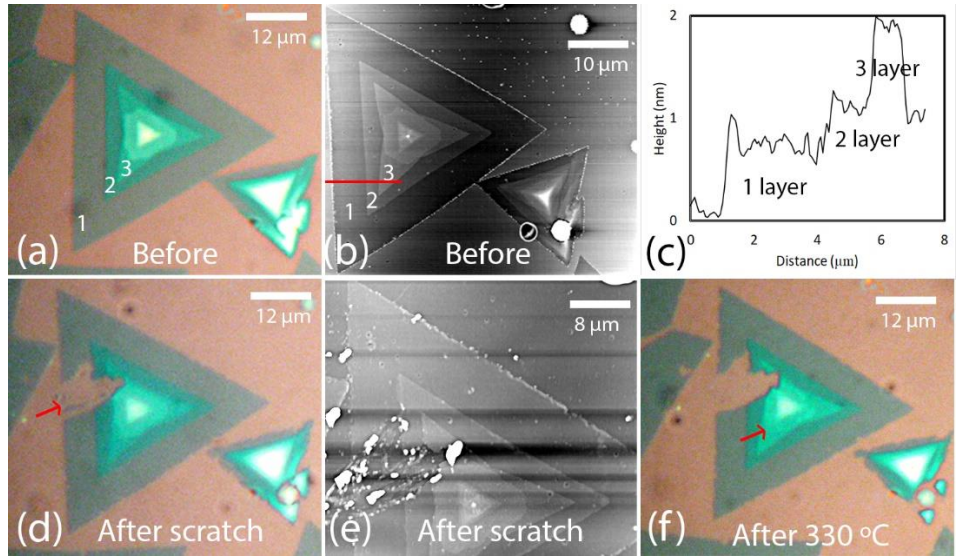

Figure 3. (a) and (b) Optical and AFM images, respectively, of an $MoS_2$ multilayer sample grown a month prior. The numbers 1, 2 and 3 denote monolayer, bilayer and trilayer, respectively. (c) AFM height profile along the red line in (b). The height of the second $MoS_2$ island is about 0.6 nm, and the height of the third $MoS_2$ island is about 0.6 nm, corresponding to a bilayer and trilayer, respectively. (d) and (e) Optical and AFM images, respectively, of the same sample after the AFM tip was used to create a scratch indicated by the arrow in (d). (f) Optical image of the $MoS_2$ sample after heating in air at 330 ºC for 2 hours. Etch pits are visible on the fourth layer, as indicated by the arrow.

To study the stability of BLs and thicker layers in a practical time span of less than a year, we put samples through the accelerated degradation process.[6] Figures 3 (a) and (b) show optical and AFM images, respectively, of an $MoS_2$ multilayer grown a month prior consisting of more than 4 layers forming a pyramidal structure. Figure 3 (c) shows an AFM height profile along the line in Figure 3 (b). The height of the second MoS$_2$ layer is about 0.6 nm, which is in agreement with the value for the ML-to-BL height difference.[23] In addition, the height of the third MoS$_2$ layer is also about 0.6 nm. This multilayer structure has the advantage of allowing a comparison of degradation, under the same conditions, as a function of number of layers. In order to investigate how BLs and thicker layers degrade given direct edge exposure to the SiO$_2$ substrate, we introduced a scratch on the left-hand side of the sample using the AFM cantilever. Figures 3 (d) and (e) show optical and AFM images, respectively, of the sample after the AFM cantilever was used to create the scratch, which is indicated by the arrow in Figure 3 (d). Figure 3 (f) shows an optical image of the MoS$_2$ structure immediately after preheating in air at 330 ºC for 2 h. While the basal plane of the BL does not appear to have been significantly affected by the preheating, the fourth layer appears to show etch pits, as indicated by the arrow in Figure 3 (f). This is consistent with previous reports that have shown that thick-layer MoS$_2$ starts to develop etch pits at lower temperatures than thin-layer MoS$_2$.[26] This behavior is in contrast to that of graphene in which the ML develops etch pits at lower temperatures than thicker layers.[27]

After preheating, the MoS$_2$ sample was left under ambient conditions. Figures 4 (a) and (b) show optical and AFM images, respectively, of the sample after 2.5 weeks of exposure to ambient conditions. There are regions of light optical contrast developing on the edges of the ML that correspond to raised dendritic regions in the AFM image. These observations are consistent with our previous report on the rapid degradation of MoS$_2$ MLs after preheating.[6] We note that dendrites are growing from the scratch edge into the ML, and have covered the entire width of the edge of the ML. However, despite the BL and thicker layers being exposed to both the substrate and the dendrites on the ML, there are no visible dendritic growths on the BL or thicker layers. Figure 4 (c) shows a high-resolution AFM image of the sample confirming the existence of etch pits on the fourth layer, as indicated by the red arrow, which was inferred from the optical image shown in Figure 3 (e). There are also less noticeable etch pits on layers thicker than four. Figures 4 (d) and (e) show optical and AFM images, respectively, of the sample after 9 weeks under ambient conditions. Figure 4 (f) shows a high-resolution AFM image showing that the etch pits have not changed, whereas the dendrites on the ML have grown across the entire width of the ML and have stopped at the ML-BL boundary. Figures 4 (g) and (h) show optical and AFM images, respectively, of the sample after 15 weeks of exposure to ambient conditions. Figure 4 (i) shows a high-resolution AFM image after 15 weeks showing that the dendrites have spread over the ML but stopped at the BL. In addition, the fourth layer containing etch pits has not significantly changed since 2.5 weeks after heating. The structural stability of preheated BLs and thicker layers was observed in the hundreds of samples and substrates that we studied, examples of which are shown in Figure S5.

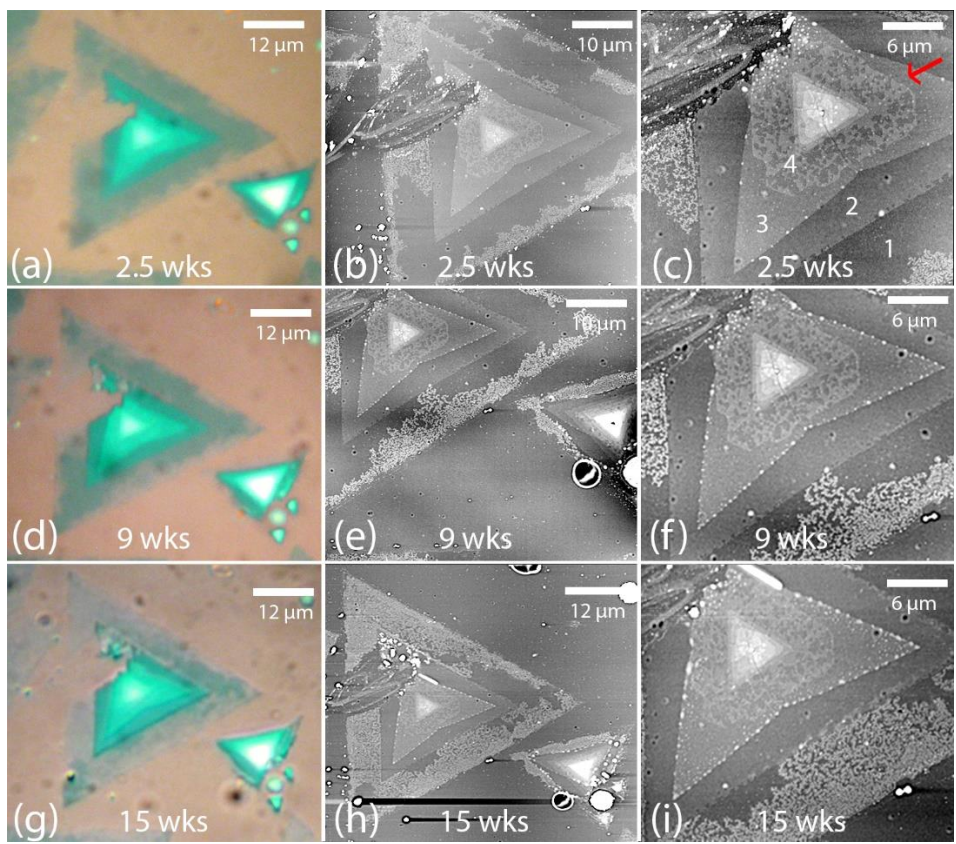

Figure 4. (a)-(c) Optical and AFM images of the MoS$_2$ multilayer sample shown in Figure 3 after 2.5 weeks of exposure to ambient conditions. Dendrites have started to develop on the monolayer, but not on the bilayer and thicker layers. The arrow in (c) indicates the fourth layer that has developed etch pits. (d-f) Optical and AFM images of the sample after 9 weeks of exposure to ambient conditions. (g-i) Optical and AFM images of the sample after 15 weeks of exposure to ambient conditions.

We also studied the sample using Raman and PL spectroscopies. Figure 5 (a) shows an AFM image of the sample after 42.5 weeks of exposure to ambient conditions. Figures 5 (b) and (c) show maps of the $E_{2g}$ and the A exciton peaks, respectively, after 42.5 weeks of exposure to ambient conditions. The same numbers in Figures 5 (a-c) indicate the same sample regions. As was the case in the sample shown in Figure 2, the ML regions with dendrites have very weak $E_{2g}$ and A exciton peaks. The ML region indicated by the number 1 has not yet degraded. This region also has a very large PL response, as shown in Figure 5 (c). The $E_{2g}$ and A exciton heights appear very uniform across the BL and thicker layers, except for several large circular regions of lower height, as indicated by the arrow in Figure 5 (b). The circular regions also appear in the AFM image after 42.5 weeks, as indicated by the arrow in Figure 5 (a). These circular regions are first observed in AFM images after 18 weeks of exposure, as shown in Figure S6. These circular regions were not observed in the AFM images in Figures 4 (h) and (i) that were taken after 15 weeks of exposure. We note from the AFM images that the size of these circular regions does not change significantly between 18 and 42.5 weeks. The appearance of these regions in less than 3 weeks and their unchanging size lead us to conclude that they are not caused by degradation, but rather by contaminants, either from ambient exposure or from deposition of material by the AFM tip. As shown in Figure S7, maps of fitted peak heights, positions and widths of the $E_{2g}$ and $A_{1g}$ peaks show that in the circular regions the $E_{2g}$ position shifts to lower wavenumbers by as much as 1.1 cm$^{-1}$ and the FWHM increases by as much as 2.3 cm$^{-1}$, while the $A_{1g}$ peak position and width remains similar to the surrounding layers. It has been reported that strain causes the $E_{2g}$ peak position to decrease and its FWHM to increase, while not significantly affecting the position or width of the $A_{1g}$ peak.[28] The contaminants may be applying strain to the $MoS_2$, causing shifts and broadening in the $E_{2g}$ peak.

To determine if the Raman and PL spectra significantly change after continued exposure to ambient conditions, we took spectra after an additional 22 weeks of exposure for a total exposure of 64.5 weeks. Figure 5 (d) shows an optical image after 64.5 weeks of exposure. Figures 5 (e) and (f) show Raman and PL spectra, respectively of the BL region taken at the location of the cross labeled by the letter "i" in Figure 5 (b) after 42.5 and 64.5 weeks of exposure. The fitted $E_{2g}$ and $A_{1g}$ peaks of both the 42.5 week and 64.5-week exposed samples are centered at about 382 cm$^{-1}$ and 405 cm$^{-1}$, respectively, and have a FWHM of 5.2 cm$^{-1}$ and 6.7 cm$^{-1}$, respectively. In addition, the B and A exciton peaks of both the 42.5 and 64.5-week-old sample are centered at around 1.94 eV and 1.80 eV, respectively. Figures 5 (g) and (h) show Raman and PL spectra, respectively of the fourth-layer region taken at the location of the cross labeled by the letters "ii" in Figure 5 (b) after 42.5 and 64.5 weeks of exposure. The fourth layer has the $E_{2g}$ and $A_{1g}$ peaks, and the B and A exciton peaks at about 381 cm$^{-1}$, 407 cm$^{-1}$, 1.90 eV, and 1.78 eV, respectively. These values are in agreement with those reported in the literature for as-grown samples.[24,25] These results provide an overview of the various effects that occur due to preheating MoS$_2$ based on the thickness of the sample. Monolayers degrade under ambient conditions, while BLs and thicker layers degrade significantly less with thicker layers more prone to develop etch pits during preheating.

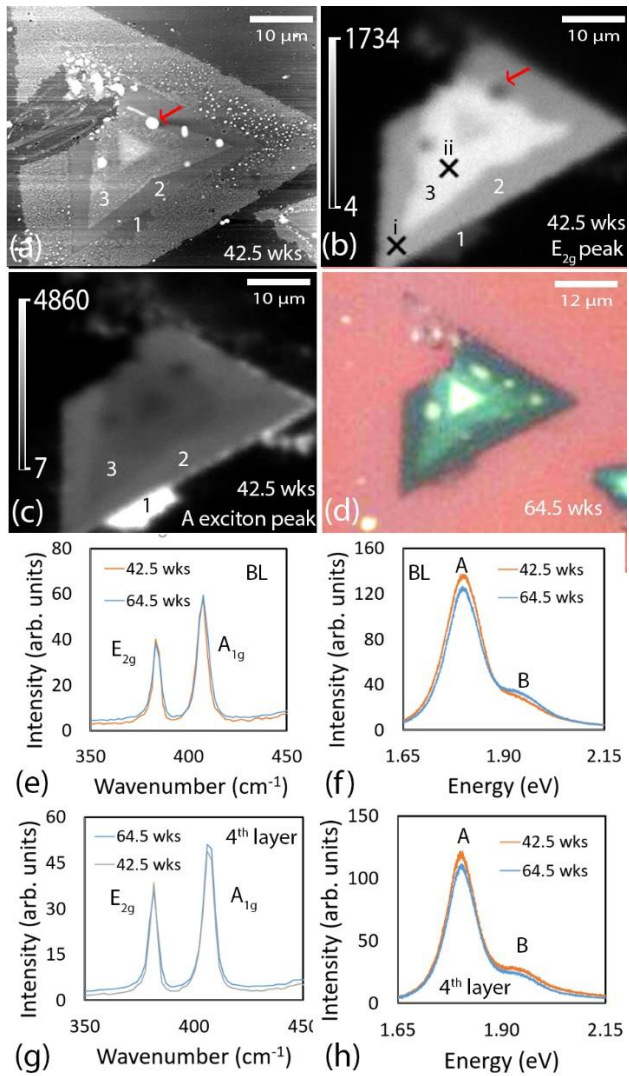

Figure 5. (a) AFM image of the multilayer sample shown in Figure 4 after 42.5 weeks of exposure to ambient conditions. The numbers 1, 2 and 3 indicate monolayer, bilayer and trilayer regions, respectively. (b) and (c) Raman and photoluminescence maps of the sample showing the relative heights of the $E_{2g}$ and the A exciton peaks, respectively. (d) Optical image of the sample after 64.5 weeks of exposure to ambient conditions. (e) and (f) Raman and photoluminescence spectra, respectively, of the bilayer taken at the location of the cross labeled "i" in (b) after 42.5 and 64.5 weeks of exposure to ambient conditions. (g) and (h) Raman and photoluminescence spectra, respectively, of the fourth layer taken at the location of the cross labeled "ii" in (b) after 42.5 and 64.5 weeks of exposure to ambient conditions.

We also studied preheated BLs with a negligible thicker multilayer on top. The objective was to determine if the thicker layers above the BL in the multilayer sample in Figure 5 contributed to its stability. Figure 6 (a) shows an optical image of the as-grown sample. The sample was then preheated at 285 °C for 2 h and exposed to ambient conditions. Figures 6 (b) and (c) show optical and AFM images, respectively, of the sample after 4 weeks of exposure. Dendrites have started growing on the ML and have stopped at the BL. Figures 6 (d) and (e) show height maps of the $E_{2g}$ and A exciton peaks, respectively. Degraded regions on the ML have lower Raman and PL heights, but the BL remains very uniform in height, even though this particular BL sample contains grain boundaries. Figure S8 shows fitted peak maps of the $E_{2g}$ and $A_{1g}$ peaks, showing very uniform position, width, and height over the BL. To study the effects of continued exposure to ambient conditions, we collected spectra after an additional 34 weeks of exposure. Figure 6 (f) shows an optical image of the sample after a total of 38 weeks in ambient conditions. Figure 6 (g) shows Raman spectra collected at the location of the black cross in Figure 6 (d) after 4 weeks and 38 weeks of air exposure. The fitted positions of the $E_{2g}$ and $A_{1g}$ peak at both 4 weeks and 38 weeks are 382 $cm^{-1}$ and 405 $cm^{-1}$, respectively. The fitted FWHM of the $E_{2g}$ and $A_{1g}$ peak at both 4 weeks and 38 weeks are about 5.1 $cm^{-1}$ and 7.0 $cm^{-1}$, respectively. Figure 6 (h) shows PL spectra collected at the same black cross, normalized to the Si Raman peak at 520 $cm^{-1}$. The peak positions of the B and A excitons of both the 4 week and 38-week exposed sample are 1.93 eV and 1.80 eV, respectively. The peaks also have very comparable heights. These values are consistent with those in the literature for as-grown samples.[24,25]

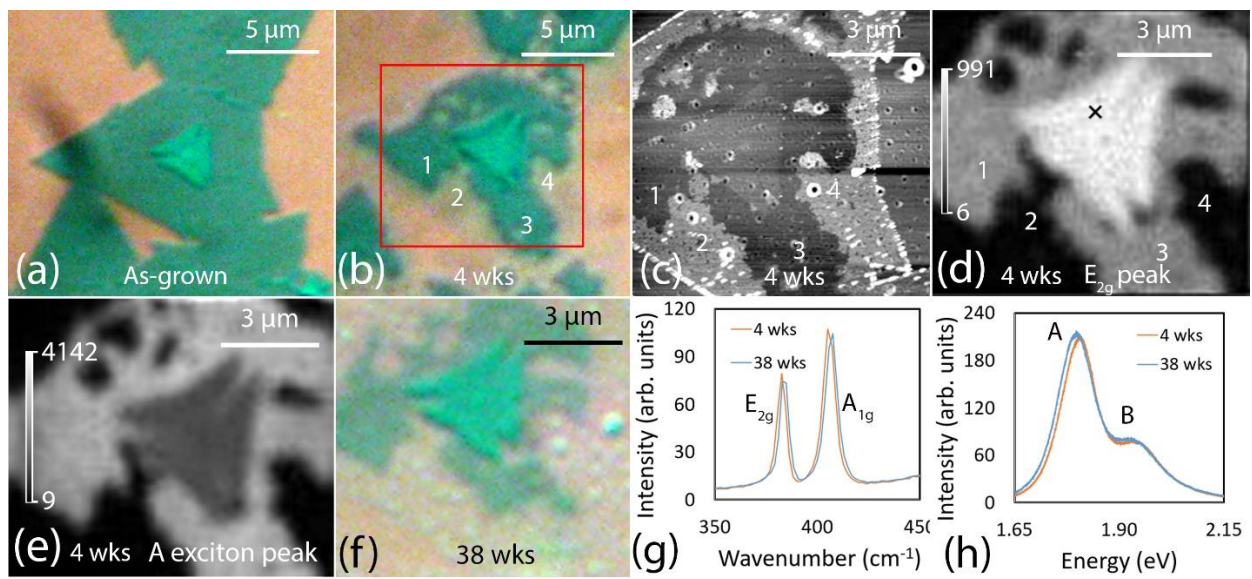

Figure 6. (a) Optical image of an $MoS_2$ bilayer 1 week after growth. (b) and (c) Optical and AFM images, respectively, of the sample after it was preheated in air at 330 °C for 2 h and then left under ambient conditions for 4 weeks. The AFM image is of the boxed region in (b). (d) and (e) Raman and photoluminescence maps of the sample, showing the relative heights of the $E_{2g}$ and the A exciton peaks, respectively. (f) Optical image of the sample after 38 weeks of exposure to ambient conditions. (g) and (h) Raman and photoluminescence spectra, respectively, of the bilayer taken at the location of the cross in (d) after 4 and 38 weeks of exposure to ambient conditions.

A mechanism for ML degradation proposed by Kotsakidis et al.[4] involves a FRET process in which photoexcited trions in the singlet state, $S_1$, act as photosensitizers that excite molecular $O_2$ to the singlet state, denoted in spectroscopic notation by $^1\Delta_g$. Singlet $O_2$ is a strong reactive oxygen species due to its electronic configuration and degrades the $MoS_2$. The steps in a typical FRET process are photoexcitation of a photosensitizer to the singlet state, $S_1$, followed by quenching of the $S_1$ state to the triplet state, $T_1$, by $O_2$ molecules in the triplet ground state, $3\Sigma_g^-$. The $T_1$ state is then quenched by $3\Sigma_g^-$ $O_2$ molecules. As a result of these processes, the $3\Sigma_g^-$ $O_2$ molecules are excited to the $^1\Delta_g$ state. A FRET process proceeds at a high rate in a direct band gap semiconductor such as ML $MoS_2$ in which radiative transitions occur at a high rate.[4] Such a mechanism was also proposed by Ding et al.[29] to explain the production of $^1\Delta_g$ $O_2$ in TMD quantum dots in solution that were exposed to visible light. These authors found a significant increase in the production of $^1\Delta_g$ $O_2$ with an increase in defects such as sulfur vacancies. This was attributed to the defects modifying the bandgap, resulting in an increase in the photosensitizing rate. An increase in $^1\Delta_g$ $O_2$ production at defects is consistent with the observation that ML degradation in ambient air starts preferentially along defect sites, such as grain boundaries and edges.[3,4,6] The FRET mechanism is consistent with our observation that BL and thicker-layer $MoS_2$ films do not degrade since these materials have indirect band gaps and consequently low radiative transition rates. The other type of mechanism proposed by Kotsakidis et al.[4] to explain ML degradation is a photocatalytic process in which photoexcited electrons and holes catalyze a hydrogen evolution reaction (HER) with adsorbed $H_2O$ and $3O_2$ molecules, similar to the HER reaction in $TiO_2$. This reaction produces various reactive oxygen species that can degrade the ML.[4] Such photocatalytic reactions are likely to occur at a faster rate in an indirect band gap semiconductor than in a direct band gap semiconductor because photoexcited electrons and holes in indirect band gap semiconductors have longer lifetimes due to the absence of direct radiative transitions. For example, in $TiO_2$ the anatase phase is significantly more photocatalytic than the rutile and brookite phases. It has recently been computationally shown that a main reason for this is that the anatase phase has long electron and hole lifetimes as a consequence of its indirect bandgap, while the rutile and brookite phases have direct band gaps.[30] Since we do not observe degradation in BL and multi-layer $MoS_2$ samples, we conclude that photocatalytic processes that may occur on the surfaces of these materials do not cause significant degradation, and consequently are not likely to cause significant degradation of MLs since in MLs such photocatalytic processes would occur at a lower rate.

Our results that BL and thicker-layer $MoS_2$ are significantly more stable than MLs is different from the results of Budania et al.[5] who reported that exfoliated multilayer $MoS_2$ flakes decompose when left in ambient air at RT and 60% RH. These authors found that the decomposition of multilayers involves the formation of speckles on the surface. Unlike the irreversible degradation of MLs, they found that the speckles reduce in size and number after vacuum annealing the samples at 200 ºC for 2 h. We conjecture that the speckles are caused by thick water films produced by condensation due to the high humidity environment. Under water, multilayer $MoS_2$ has been reported to decompose by a different mechanism involving the dissolution of oxides in water. The removal of the oxides leaves the surface susceptible to continued oxidation.[31–33] For example, Zhang et al.[32] studied the oxidation of thick $MoS_2$ sheets placed in water for up to 12 h and observed etching of the basal plane that was attributed to dissolution of oxides. Also, Parzinger et al.[33] attributed the photocorrosion of exfoliated ML and trilayer $MoS_2$ in water to dissolution of oxides. We propose that in ambient air at humidities of about 43%, water films that dissolve oxides do not form on the multilayer surface, and the oxidized sites protect the surface from further oxidation. In MLs, degradation occurs by reaction with singlet oxygen produced by the FRET mechanism.

**Conclusions**

CVD-grown BL and thicker-layer $MoS_2$ films do not significantly degrade under ambient conditions at RT, 43% RH and incident room light intensity of 1000 Lux for the period tested of up to 2 years for non-preheated films and 64.5 weeks for preheated films. The structural and optical properties of aged BL and thicker-layer films were similar to as-grown films as determined by AFM, and Raman and PL spectroscopies. This contrasts with MLs that almost completely degrade during this time. These results support a previously reported mechanism in which ML TMD degradation is due to a FRET mechanism, since such a mechanism would occur at a higher rate in a direct band gap semiconductor than in an indirect band gap semiconductor. These results also show that photocatalytic processes do not play a significant role in ML degradation, since such processes would occur at a higher rate in an indirect band gap semiconductor. We expect that other TMDs in which the ML has a direct band gap and the BL and thicker layers have indirect band gaps will also have this property. Our results should motivate research into BL and thicker-layer TMDs since these materials are likely to be significantly more stable under ambient conditions.

**Materials and Methods**

The $MoS_2$ samples were grown using CVD on 300 nm thick $SiO_2$ layers on Si using the same procedure we have previously reported.[34] The $SiO_2$/Si substrate and crucibles containing S and $MoO_3$ were placed in a tube furnace. The tube furnace was then flushed at RT with ultrahigh purity Ar at a flow rate of 500 sccm for 10 min. Growth was carried out by ramping the temperature of the tube furnace at 25 °C/min up to 800 °C with an Ar flow rate of 100 sccm. The temperature was kept at 800 °C for 20 min, after which the tube furnace was turned off and allowed to cool to RT. We used a NaCl precursor during the growth, which has been reported to contribute to the formation of multilayer $MoS_2$.[13] We have previously reported that the NaCl precursor does not contribute to ML degradation.[6] The samples were preheated in a different tube furnace in air at atmospheric pressure. Raman and PL spectroscopy maps were obtained using a Renishaw inVia Raman Microscope with a 532 nm laser and Renishaw Centrus CCD detector, laser power of 0.5 mW, spot size of 764 nm, grating of 1800 lines/mm, and a spectral resolution of 1.75 cm$^{-1}$. The maps were acquired with a stage step size of 0.1 µm and integration time of 0.1 s per point. The AFM system used was an Ambios Q-scope.

**Acknowledgements**

We would like to acknowledge Dave Banerjee, Aryan Agarwal, Ernest Lu, and Mihir Khare for assistance and useful discussions. We also thank Guido Verbeck for the use of facilities. This work was supported by an Emerging Frontiers in Research and Innovation (EFRI) grant from the National Science Foundation (NSF Grant No. 1741677). KY acknowledges the support from the NSF Summer REM program at UNT. The authors declare no competing interest.

**Author Contributions**

J.F.O first noticed the structural degradation from AFM measurements. He conducted all the AFM characterization and wrote the first draft of the manuscript. E.H. and J.C. provided facilities for and conducted the majority of the Raman spectroscopy. N.K.G. and W.Z. also collected Raman data. Y.J. and U.P. provided facilities for and grew CVD MoS2 samples. K.Y. first observed degradation of $MoS_2$ in optical microscopy, and he monitored and took images of the samples with contributions from I.O. B.S. and A.N. provided useful discussions. K.Y. and J.P. conceived of experiments, analyzed the data, made the figures, and wrote the manuscript. The manuscript was edited through contributions from all the authors.

**Supporting Information**

Supporting Information Available: Additional optical images, AFM images, and fitted Raman and photoluminescence maps. This material is available free of charge via the Internet at http://pubs.acs.org.

# *Supporting Information*
# Long-term Stability of Bilayer MoS$_2$ in Ambient Air

John Femi-Oyetoro[1], Kevin Yao[1, a], Evan Hathaway[1], Yan Jiang[1], Ibikunle Ojo[1], Brian Squires[1], Arup Neogi[1], Jingbiao Cui[1], Usha Philipose[1], Nithish Kumar Gadiyaram[2], Weidong Zhou[2], and Jose Perez[1, *]

[1]Department of Physics, University of North Texas, Denton, TX 76203, United States

[2]Department of Electrical Engineering, University of Texas at Arlington, Arlington, Texas 76019, United States

*Corresponding author.

E-mail address: jperez@unt.edu (J.M. Perez).

[a]Current address: Texas A&M University, College Station, TX 77843, United States

1. Optical image of bilayer stability without sample preheating

We observe the stability of bilayers (BLs) in all of our naturally aged samples, regardless of location on the substrate, growth batch, or time left in air up to 2 years. Figure S1 shows representative optical images of a 1.5-year-old CVD-grown $MoS_2$ film on an $SiO_2$ substrate. In Figure S1 (a), the arrows indicate $MoS_2$ islands where the monolayer (ML) has developed regions of light optical contrast, indicating degradation,[1] but the BL appears optically uniform, suggesting that it has not degraded. Figure S1 (b) is a high-magnification optical image of the boxed region in Figure S1 (a) showing an AB-stacked BL that has not degraded. Figures S1 (c) shows an optical image of a different area of the same $SiO_2$ substrate. The arrows indicate $MoS_2$ islands where the ML has developed regions of light optical contrast, but the BL appears optically uniform. Figure S1 (d) is a high-magnification optical image of the boxed region in Figure S1 (c) showing an AA-stacked BL that has not degraded.

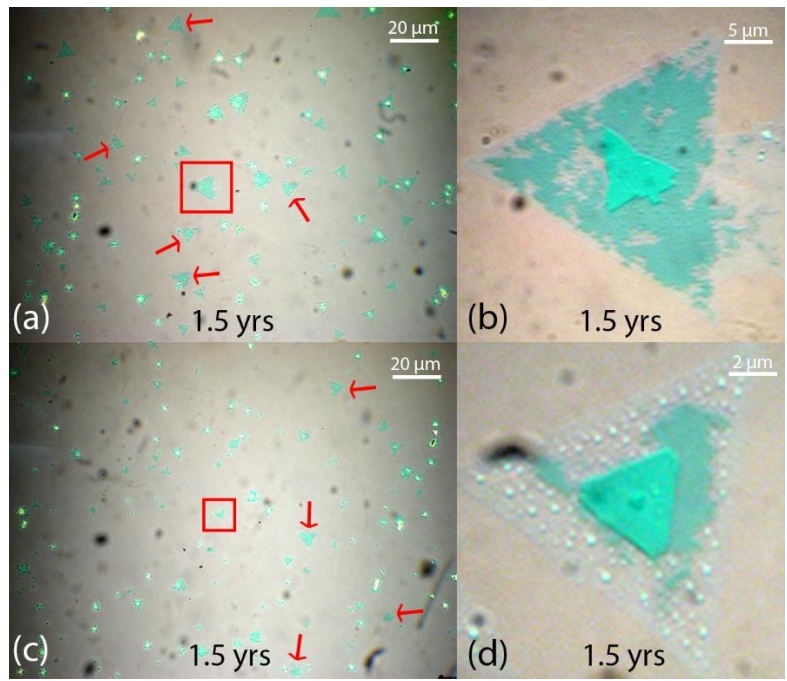

Figure S SEQ Figure \* ARABIC 1. (a) Low-magnification optical image of an $MoS_2$ sample that has been left under ambient conditions for 1.5 years. The red arrows point at specific examples where the degradation is proceeding over the monolayer but the bilayer remains stable. (b) High-magnification optical image of the boxed area in (a) showing an AB-stacked bilayer. (c) Wide-area optical image of a different sample region in which arrows indicate examples of degradation. (d) High-magnification optical image of the boxed area in (c) showing an AA-stacked bilayer.

2. Irreversibility of degradation

The formation of dendrites in ML $MoS_2$ was found to be irreversible with respect to heating in vacuum. Figures S2 (a) and (b) show optical and AFM images, respectively, of a CVD-grown $MoS_2$ sample that has been preheated in air at 290 °C for 2 h and then left in ambient conditions for 4 weeks. The arrows in Figure S2 (b) show regions of elevated dendrites, indicating degradation.[1,2] The sample was then annealed in a vacuum at about $10^{-5}$ Torr at a temperature of 500 °C for 2 h. Figure S2 (c) and (d) show optical and AFM images, respectively, of the sample immediately after the vacuum anneal. The arrows in Figure S2 (d) indicate the same regions of elevated dendrites as in Figure S2 (b), showing that the degradation remained.

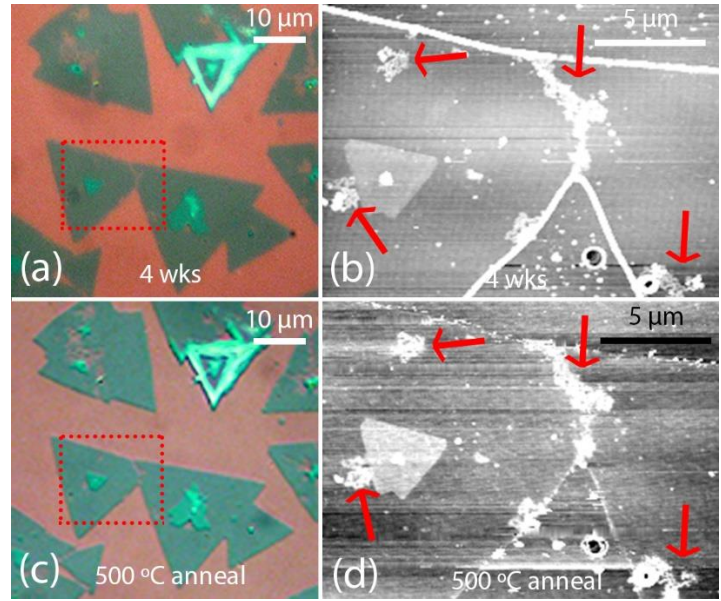

Figure S SEQ Figure \* ARABIC 2. (a) Optical image of an MoS$_2$ sample preheated at 290 °C for 2h in air and then left under ambient conditions for 4 weeks. (b) AFM image of the boxed area in (a). Dendrites are visible, as indicated by the arrows. (c) Optical image of the sample after vacuum annealing at 10-5 Torr and 500 °C for 2 h. (d) AFM image of the boxed area in (c). The dendrites are still visible, as shown by the arrows, indicating that the degradation has not been removed.

3. 1.5-year-old non-preheated sample: Fitted Raman height, position, and width maps, and PL map

We used the Renishaw Wire software (Renishaw WiRE 5.2 software, Renishaw) to generate maps of the fitted peak heights, positions, and FWHMs of the Raman $E_{2g}$ and $A_{1g}$ peaks. Such maps are useful in determining slight variations in peak position and FWHM. For example, sample strain has been shown to produce a shift in position and broadening of the $E_{2g}$ peak without a corresponding shift and broadening in the $A_{1g}$ peak.[3] Figure S3 shows fitted peak maps of the BL sample shown in Figure 2 of the paper after 1.5 years of exposure to ambient conditions. Figures S3 (a-c) show maps of the fitted height, position, and FWHM of the $E_{2g}$ peak, respectively, and Figures S3 (d-f) show maps of the fitted height, position, and FWHM of the $A_{1g}$ peak, respectively. All of the fitted peak maps were generated from the spectra collected during the run for the map shown in Figure 2. Figures S3 (a) and (d) show that the BL has a large $E_{2g}$ and $A_{1g}$ peak height, as expected, and the heights are relatively uniform throughout the BL. On the other hand, the ML regions contain areas of significantly decreased height, which represents degradation. Labels 1-3 in (a)-(f) serve as a visual guide to indicate the same degraded regions. We note that the position and width of the $E_{2g}$ and $A_{1g}$ peaks are also uniform on the BL. On the ML, they exhibit slight shifts towards higher wavenumbers and broadening when closer to the degraded regions. The $E_{2g}$ shows a more significant peak shift and broadening compared to the $A_{1g}$ peak, as has been previously reported for ML degradation.[1]

Figure S4 (a) shows the PL height map copied from Figure 2 (c) of the paper for convenience. Figure S4 (b) shows orange and blue PL spectra collected at the orange and blue crosses, respectively, shown in Figure S4 (a). The orange cross is located in a region where the ML has not significantly degraded, and the blue cross is located in a region where the ML has significantly degraded. In Figure S4 (b), the A exciton height is decreased by a factor of about 4, and shifts to higher energy by about 0.5 eV, consistent with previous reports [2]. We observe a decrease in the A exciton height by up to a factor of 20 in degraded areas.

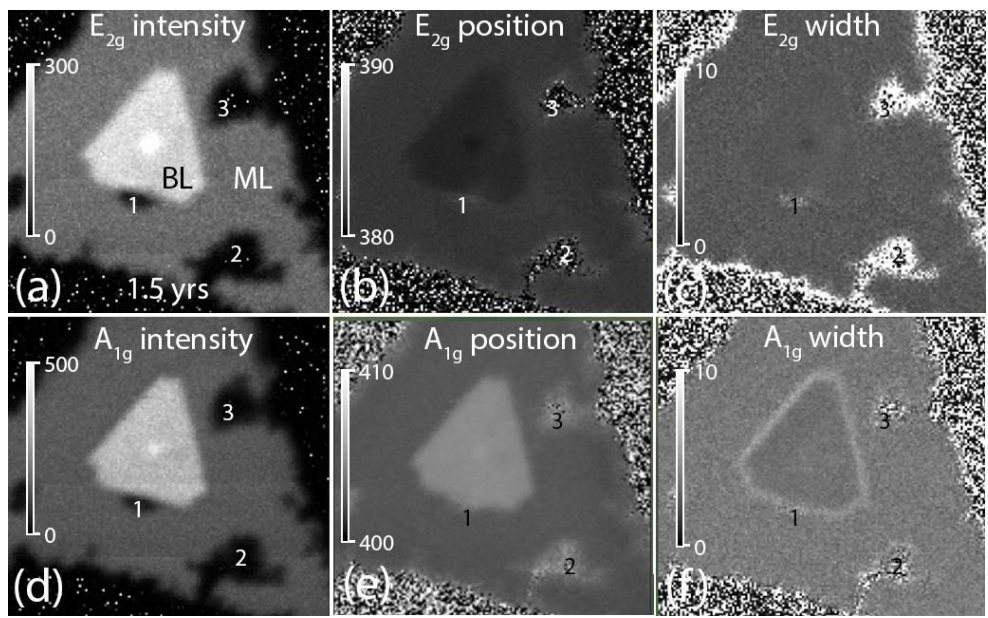

Figure S3. Fitted peak maps of the 1.5-yr-old sample from Figure 2 in the paper. The numbers 1, 2 and 3 indicate regions of degradation on the monolayer. (a-c) Peak height, position, and FWHM maps, respectively, of the $E_{2g}$ peak. (d-f) Peak height, position, and FWHM maps, respectively, of the $A_{1g}$ peak.

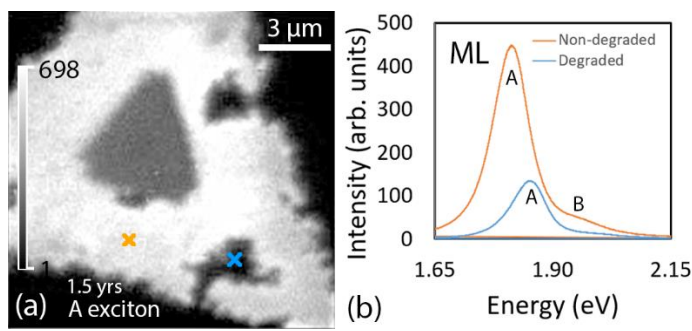

Figure S4. (a) Raman height map of the A exciton peak of the 1.5-yr-old sample, copied from Figure 2 in the paper. (b) PL spectra taken at the locations indicated by the blue and orange crosses in (a). The blue and orange curves were taken at the locations indicated by the blue and orange crosses, respectively.

4. Optical image of bilayer stability with sample preheating

Figure S5 shows representative optical images of $MoS_2$ samples that were preheated in air at 330 °C for 2h, showing that BLs are also stable in ambient conditions after preheating. Preheating accelerates ML degradation.[1] Figure S5 (a) shows an optical image of the sample taken immediately after preheating. The BLs indicated by numbers 1 and 3 are AB stacked, and the BL indicated by the number 2 is AA stacked. Figure S5 (b) shows a high-magnification optical image of the boxed region in (a). Figures S5 (c) and (d) show optical images of the same sample areas as in (a) and (b), respectively, after the sample was left under ambient conditions for 9 weeks. We have examined hundreds of preheated samples and not observed BL degradation.

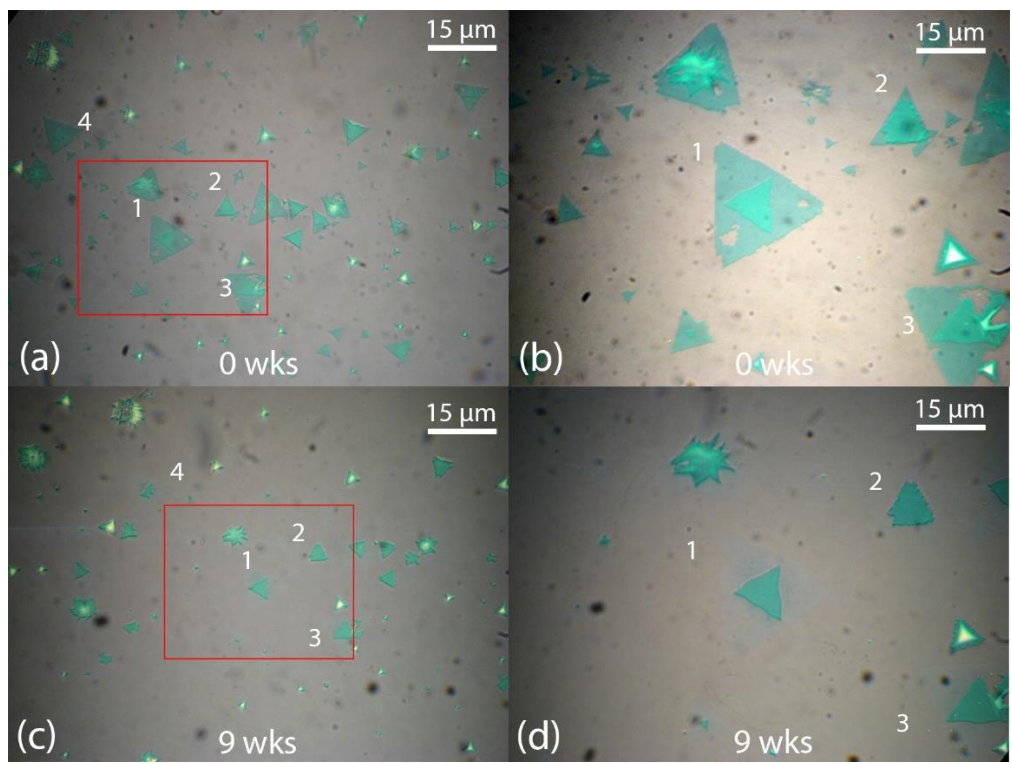

Figure S5. (a) Optical images of an MoS$_2$ sample immediately after heating in air at 330 ºC for 2 h. The numbers 1, 2 and 3 indicate AB, AA and AB-stacked bilayers, respectively. (b) High-magnification optical image of the boxed area in (a). (c) Optical image of the same area in (a) after 9 weeks of exposure to ambient conditions. (d) High-magnification optical image of the boxed area in (c). The monolayers have completely degraded whereas the bilayers remain visible with no sign of degradation.

5. Multilayer sample: Optical and AFM images after 18 weeks, and fitted Raman height, position, and width maps after 42.5 weeks

Figures S6 (a) and (b) show optical and AFM images, respectively, of the multilayer MoS$_2$ sample shown in Figures 4 and 5 of the paper after 18 weeks of exposure to ambient conditions. The red arrow indicates the bright circular region that was also observed after 42.5 weeks of exposure in Figure 5 (a) of the paper. The bright circular region has remained almost unchanged in shape and size between 18 weeks and 42.5 weeks. We note that we do not observe this region after 15 weeks, as shown in Figures 4 (g) and (h) of the paper. This region appeared over a period of at most 3 weeks and then did not significantly change. We conclude it is not a product of degradation and may be contamination deposited on the sample.

Figure S7 shows fitted peak maps preheated multilayer sample shown in Figure 5 of the paper after 42.5 weeks of exposure to ambient conditions. The fitted maps were generated from the spectra collected during the run for the map shown in Figure 5 (b) of the paper. Figures S7 (a-c) show maps of the fitted height, position, and width of the E$_{2g}$ peak, respectively, and Figures S7 (d-f) show maps of the fitted height, position, and width of the A$_{1g}$ peak, respectively. The E$_{2g}$ and A$_{1g}$ peaks of the BL and thicker layers have uniform heights, positions, and widths with the exception of areas indicated by the arrows in Figure S7, which correspond to the large circular regions observed in the AFM image shown in Figure 5 (a) of the paper. These circular regions exhibit lower E$_{2g}$ and A$_{1g}$ heights. We note that in the circular regions, Figure S7 (b) shows that the E$_{2g}$ peak shifts to lower wavenumbers by as much as 1.1 cm$^{-1}$, while Figure S7 (e) shows that the position of the $A_{1g}$ peak does not noticeably shift from that of the surrounding areas. In addition, Figure S7 (c) shows that the width of the $E_{2g}$ peak in the circular regions increases by as much as 2.3 cm$^{-1}$, while Figure S7 (f) shows that the width of the $A_{1g}$ peak does not change significantly. Sample strain has been shown to result in blueshifting and broadening of the $E_{2g}$ peak with no significant change in the position and width of the $A_{1g}$ peak.[3] Since both the $E_{2g}$ and $A_{1g}$ peaks exhibit decreased height, we believe that a contaminant may have been deposited on the sample, which is blocking the $E_{2g}$ and $A_{1g}$ height signals and applying strain to the underlying $MoS_2$ lattice.

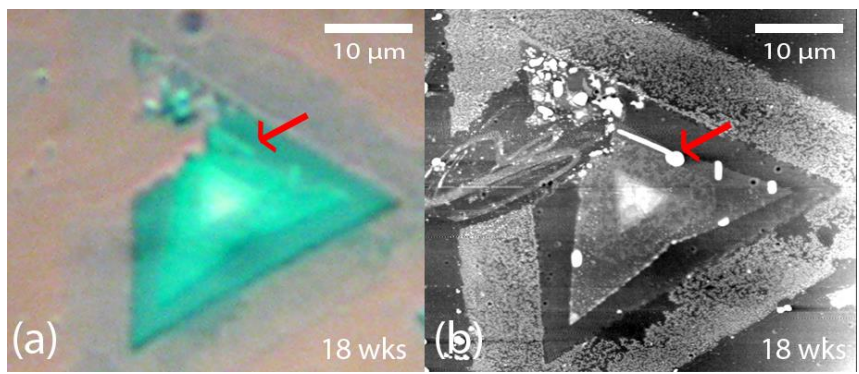

Figure S6. (a) and (b) Optical and AFM images, respectively, of the $MoS_2$ sample shown in Figure 4 of the paper after 18 weeks of exposure to ambient conditions. The arrows indicate the location of bright circular regions seen in Figure 5 (a) of the paper.

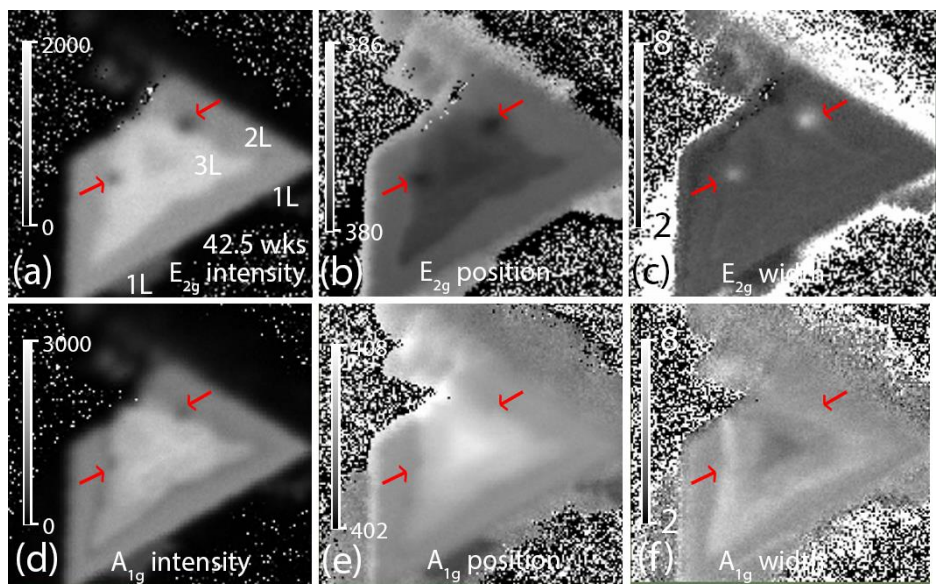

Figure S7. Fitted peak maps of the multilayer sample shown in Figure 5 (a) of the paper after 42.5 weeks of exposure to ambient conditions. (a-c) Peak height, position, and FWHM maps, respectively, of the $E_{2g}$ peak. (d-f) Peak height, position, and FWHM maps, respectively, of the $A_{1g}$ peak.

6. 4-week-old preheated sample: Fitted Raman height, position, and width maps.

Figure S8 shows fitted peak maps of the preheated BL sample shown in Figure 6 (b) of the paper that was exposed to ambient conditions for 4 weeks. The fitted maps were generated from the spectra collected during the run for the map shown in Figure 6 of the paper. Figures S8 (a-c) show maps of the fitted height, position, and FWHM of the $E_{2g}$ peak, respectively, and Figures S8 (d-f) show maps of the fitted height, position, and FWHM of the $A_{1g}$ peak, respectively. Both the $E_{2g}$ and $A_{1g}$ peaks of degraded ML regions show lower heights, shifts to higher wavenumbers, and broadening of the widths. The BL regions appear uniform.

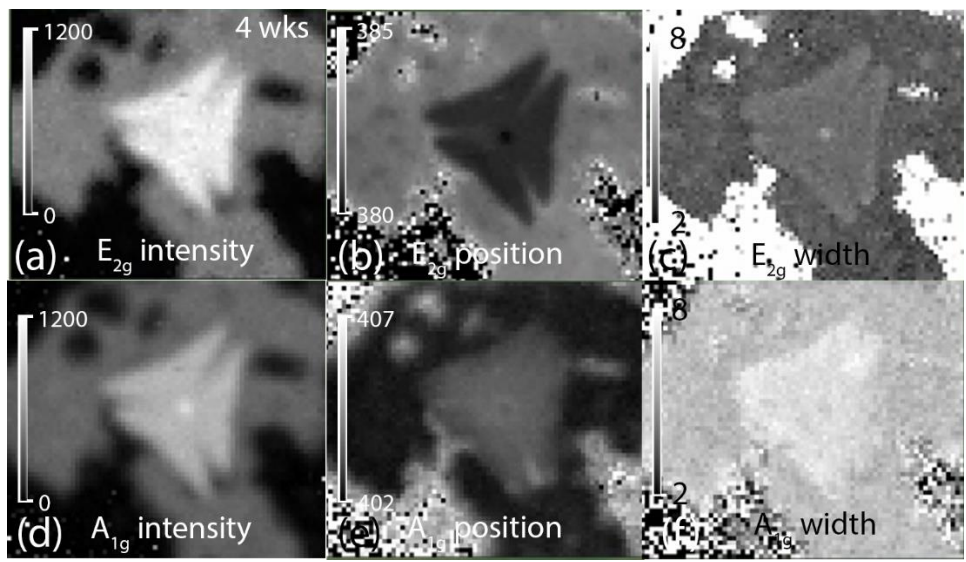

Figure S8. Fitted peak maps of the sample shown in Figure 6 (b) of the paper after 4 weeks of exposure to ambient conditions. (a-c) Peak height, position, and FWHM maps, respectively, of the $E_{2g}$ peak. (d-f) Peak height, position, and FWHM maps, respectively, of the $A_{1g}$ peak.